\documentclass[12pt,journal,draft,onecolumn]{IEEEtran}
\usepackage{cite}
\usepackage[final]{graphicx}
\usepackage{amsmath}

\def\cM{{\mathcal M}}
\def\cX{{\mathcal X}}
\def\cY{{\mathcal Y}}
\def\cP{{\mathcal P}}

\newtheorem{lemma}{Lemma}\def\lemref#1{Lemma~\ref{#1}}
\newtheorem{theorem}{Theorem}\def\thref#1{Theorem~\ref{#1}}

\def\figref#1{Fig.~\ref{#1}}
\def\bydef{:=}
\def\corr#1#2{\phi_{#1,#2}}
\DeclareMathOperator{\E}{E}
\def\apref#1{Appendix~\ref{#1}}
\def\secref#1{Section~\ref{#1}}

\normalsize

\begin{document}
\title{On the Rate Achievable for Gaussian Relay Channels Using Superposition
Forwarding}

\author{Neevan~Ramalingam and Zhengdao~Wang
\thanks{The authors are with the department of Electrical and
Computer Engineering, Iowa State University, Ames, IA, 50014 USA} }

\markboth{}{}

\maketitle

\begin{abstract}
We analyze the achievable rate of the superposition of block Markov encoding
(decode-and-forward) and side information encoding (compress-and-forward) for
the three-node Gaussian relay channel. It is generally believed that the
superposition can out perform decode-and-forward or compress-and-forward due
to its generality. We prove that within the class of Gaussian distributions,
this is not the case: the superposition scheme only achieves a rate that is
equal to the maximum of the rates achieved by decode-and-forward or
compress-and-forward individually. We also present a superposition scheme that
combines broadcast with decode-and-forward, which even though does not achieve
a higher rate than decode-and-forward, provides us the insight to the main
result mentioned above.
\end{abstract}


\IEEEpeerreviewmaketitle

\section{Introduction} \IEEEPARstart{T}{he} relay channel, introduced by van
der Meulen \cite{vand71} is a fundamental building block in network
information theory. It consists of a relay terminal assisting communication
between a source terminal and a destination terminal, facilitating a higher
data rate than that of a point to point channel. Cover and El Gamal
\cite{coga79} introduced two new coding strategies and a cut-set upper bound
for the relay channel. They derived the capacity of the degraded and reversely
degraded relay channels. Capacity results have been derived for special cases
of the relay channel like the semi-deterministic case \cite{gaar82} but the
capacity of the general relay channel is still unknown.

The main achievability strategies known for the relay channel are Decode and
Forward (DF) and Compress and Forward (CF) \cite{coga79}. The DF scheme is
also known as the general block Markov encoding scheme. The relay decodes the
transmitted message and jointly transmits the message from the source to the
destination terminal. The DF strategy is optimal and achieves the cut-set
bound when the source to relay link channel is strong. The CF scheme is known
as the side-information encoding scheme. The relay compresses the received
signal without decoding and transmits to the destination terminal. The
destination terminal treats the compressed information as side information and
decodes the original message. The CF scheme is asymptotically optimum and
achieves the cut-set bound when the relay to destination link channel is
strong, so that the received signal at the relay can be conveyed faithfully to
the destination. A combination of the two strategies that superimposes DF and
CF was also proposed in \cite[Theorem 7]{coga79}. Hereafter we refer to this
scheme as the superposition forwarding (SF). The SF scheme achieves the
capacity for the special cases of degraded, reversely degraded and
semi-deterministic relay channels. Due to the generality of the result in
\cite[Theorem 7]{coga79}, it is expected it can offer higher achievable rates
than DF or CF alone.

In this paper, we investigate the coding scheme for the general Gaussian relay
channel. The initial motivation for the work was to develop new coding
strategies with higher achievable rates. A new coding strategy was designed
which superimposes Decode and Forward and Broadcast, as presented in
\secref{sec.super}. The scheme unfortunately yields a rate that is inferior to
DF. This attempt, though not successful, prompted us to investigate the
general superiority of SF, especially for the Gaussian relay channel. It is
found that for Gaussian relay channel, within the class of Gaussian
distributions, the SF can achieve at most the larger rate achievable by DF or
CF alone --- there is no need to do superposition for Gaussian distributions
(\secref{sec.main}). We also provide one numerical example that verified the
theoretical result in \secref{sec.num}. \secref{sec.conc} concludes the paper.

Notation: For random variables $X,Y,Z$, we use $p(x,y,z)$ to denote the joint
distribution, when there is no confusion, as a short cut to
$p_{X,Y,Z}(x,y,z)$. When $X$ and $Z$ are conditionally independent given $Y$
(i.e., $X,Y$, and $Z$ form a Markov chain), we write $X-Y-Z$.

\section{Preliminaries} We present the mathematical models for the
discrete-memoryless and Gaussian relay channels in this section, and also
include the known results on achievable rates that will be used later.

\subsection{Discrete memoryless relay channel}

The general discrete memoryless relay channel (DMRC) is the same as defined in
\cite{coga79}. A brief description is given here for completeness. The DMRC is
denoted by $( \cX_1\times \cX_2,p(y_2,y_3|x_1,x_2)$, $\cY_2\times\cY_3)$,
where $\cX_1,\cX_2,\cY_2,\cY_3$ are finite sets and $p(.,.|x_1,x_2)$ is a
collection of probability distributions on $\cY_2\times\cY_3$, one for each
$(x_1,x_2) \in \cX_1\times\cX_2$; $x_1$ and $x_2$ are the transmitted symbols
at the source and the relay respectively; $y_2$ and $y_3$ are the received
symbols at the relay and the destination terminal.

An $(M,n)$ code for the relay channel consists of a set of integers $\cM =
\{1,2,\ldots,M \}$, an encoding function $ x_1 : \cM \rightarrow \cX_1^n $ a
set of relay functions $\{f_i\}_{i=1}^n $ such that
\[
x_{2i} = f_i \left(Y_{21},Y_{22},\ldots,Y(2i-1) \right),
\hspace{10pt} 1 \le i\le n,
\]
and a decoding function $ g: \cY_3^n \rightarrow \cM $. The joint probability
mass function on $\cM\times\cX_1^n\times\cX_2^n\times\cY_2^n\times\cY_3^n$ is
\begin{align}
p(w,x_1,x_2,y_2,y_3) = p(w)\prod_{i=1}^n p(x_{1i}|w)
	p(x_{2i}|y_{21},y_{22},\ldots,y_{2i-1}) p(y_{2i},y_{3i}|x_{1i},x_{2i}).
\end{align}
\\ \noindent Define $\lambda(w) = p(g(Y) \neq w)$ as the probability of error
of the decoding function of the relay channel and let $\lambda_n$ be the
maximal probability of error over all possible messages $w$. The rate $R =
(1/n) \log M$ of an $(M,n)$ code is said to be achievable by a relay channel
if for any $\epsilon > 0$ and for sufficiently large $n$, there exists a code
with $M \ge 2^{nR}$ such that $\lambda_n < \epsilon$.

\subsection{Gaussian relay channel}

\begin{figure}
\centering
\includegraphics[width=3.5in]{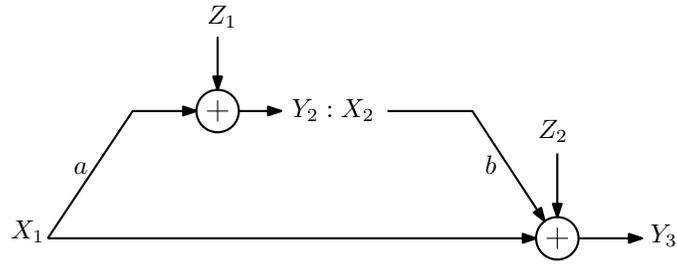}
\caption{Gaussian relay channel}
\label{figrelay}
\end{figure}

\figref{figrelay} shows the Gaussian relay channel model that we will be
using. The received symbols at the relay and the destination terminal are
given respectively by
\begin{eqnarray}
Y_2 &=& a X_1 + Z_1 \\ Y_3 &=& X_1+b X_2 + Z_2
\end{eqnarray}
where the noise terms $Z_1$ and $Z_2$ are uncorrelated zero mean Gaussian
random variables with variances $N_1$ and $N_2$ respectively, and $a$ and $b$
are the channel gain constants. As a result, we have
\begin{equation}\label{eq.ch}
	p(y_2, y_3|x_1, x_2)=\frac 1{2\pi \sqrt{N_1N_2}}
		\exp\left[ - \frac {(y_2-ax_1)^2}{2N_1}-
			\frac {(y_3-x_1-bx_2)^2}{2N_2} \right],
\end{equation}
which will be the channel assumed throughout the paper.

The average power constraints at the transmitters are
\begin{align}
\frac{1}{n} \sum_{i=1}^n x_{1i}^2(k) &\le P_1, \quad \forall k \in \cM,
\intertext{and}
\frac{1}{n} \sum_{i=1}^n x_{2i}^2 & \le P_2,\quad \forall y_2^n \in \Re^n.
\end{align}

\subsection{Known achievable rates}

We briefly review the known results in for DF, CF, and the SF. For DMRC, the
DF scheme achieves any rate less than \cite[Theorem~1]{coga79}
\begin{equation}
R_{DF} = \sup \min \{ I(X_1;Y_2|X_2),I(X_1,X_2;Y_3) \}
\end{equation}
where the supremum is taken over all possible $p(x_1, x_2)$. The CF scheme
achieves any rate less than \cite[Theorem~6]{coga79}
\begin{equation} \label{eq.RCF}
R_{CF} = \sup I(X_1;\hat{Y}_2,Y_3|X_2) ,\quad
  \text{such that } I(X_2;Y_3) \ge I(\hat{Y}_2;Y_2|X_2,Y_3)
\end{equation}
where supremum is taken over all joint probability distributions of the form
\begin{equation} \label{eq.p6}
p(x_1,x_2,y_2,y_3,\hat{y_2}) =
	p(x_1)p(x_2)p(y_2,y_3|x_1,x_2)p(\hat{y}_2|y_1,x_2).
\end{equation}
El Gamal, Mohseni, and Zahedi \cite{elmz06} put forth an equivalent
characterization of the CF scheme. That is, it achieves any rate less than
\begin{equation}\label{eq.CF2}
R_{CF} = \sup \min \{ I(X_1;\hat{Y}_2,Y_3|X_2), I(X_1,X_2;Y_3)- I(\hat{Y}_2;Y_2|X_1,X_2,Y_3) \} \\
\end{equation}
where supremum is still taken over all joint probability distributions of the
same form as in \eqref{eq.p6}. The supremum of rates achievable by
superimposing DF and CF \cite[Theorem 7]{coga79} is
\begin{multline}
R_{SF}=\sup (\min \{I(X_1;Y_3,\hat Y_2'|X_2,U)+I(U;Y_2|X_2,V),\\ \quad
I(X_1,X_2;Y_3)-I(\hat Y_2';Y_2|U,X_1,X_2,Y_3)\} ) \label{eq.th7}
\end{multline}
where the supremum is over all joint probability distributions of the form
\begin{equation}\label{eq.p7}
p(u,v,x_1,x_2,y_2',y_3,\hat{y_2}) = p(v)p(u|v)p(x_1|u)
p(x_2|v)p(y_2,y_3|x_1,x_2)p(\hat{y}_2'|x_2,y_2,u)
\end{equation}
subject to the constraint
\begin{equation}\label{eq.th7con}
I(X_2;Y_3|V) \ge I(\hat{Y}'_2;Y_2|X_2,Y_3,U).
\end{equation}
Finally, the rate is upper bounded by the cut-set bound
\begin{equation}\label{eq.cutset}
	R_{CS}=\sup \min\{I(X_1, X_2; Y_3), I(X_1; Y_2, Y_3)\},
\end{equation}
where the supremum is taken over all possible distributions $p(x_1, x_2)$.

\section{Broadcast over Decode and Forward}\label{sec.super}

Before investigating the coding scheme that superimposes CF and DF for the
Gaussian relay channel, we will first look at a simpler coding scheme. In this
scheme, partial information is decoded first at both the relay and the
destination terminals like in a broadcast channel. The remaining message is
decoded and forwarded given the partial information available at the relay and
destination terminal. The coding scheme is equivalent to superimposing
broadcast over decode and forward.

We split the message $W$ into two parts $W'$ and $W''$ with respective rates
$R'$ and $R''$. We demand $W'$ be decoded at both relay and destination. The
relay also decodes the message $W''$ which the destination could not decode
and sends this extra information to the destination in a block Markov encoding
fashion. This strategy can be designed using an auxiliary random variable $U$
and a block Markov superposition encoding explained below.

\begin{theorem}\label{th.sup}
For any relay channel ($\cX_1 \times \cX_2, p(y_2,y_3|x_1,x_2),\cY_2 \times
\cY_3$), the rate $R$ is achievable where
\begin{equation}\label{eq.bd}
R < \sup_P \{ \min\{I(U;Y_3), I(U; Y_2| X_2)\} +
	\min\{I(X_1;Y_2|X_2,U),I(X_1,X_2;Y_3|U) \}  \}
\end{equation}
and the supremum is taken over all probability distribution functions of the
form
\[
p(u,x_1,x_2,y_2,y_3) = p(u)p(x_2)p(x_1|x_2,u)p(y_2,y_3|x_1,x_2).
\]
\end{theorem}
\begin{IEEEproof}[Proof of Theorem 1]
\paragraph{Codebook Generation} Encoding is performed in $K+1$ blocks. For
each block $k$, generate $2^{nR'}$ codewords $u_k^n(s), s=1,2,\ldots,2^{nR'}$
by choosing the $u_{ki}(s)$ independently using the distribution $P_U(\cdot)$.
Generate $2^{nR''}$ codewords $x_{2k}^n(t), t = 1,2,\ldots,2^{nR''}$ by
choosing $x_{2ki}(t)$ independently using the probability distribution
$P_{X_2}(\cdot)$. Now use superposition coding and generate $2^{nR''}$
codewords $x_{1k}^n(r|s,t)$, $r=1,2,\ldots, 2^{nR''}$ for every pair of
$(u_k^n(s), x^n_{2k}(t))$, by choosing the $x_{1k,i}(r|s,t)$ independently
using $P_{(X_1|X_2,U)}(.|u_{k,i}(s),x_{2k,i}(t))$.

\paragraph{Encoding} Let $s_k$ be the message index of $W'$ and $t_k$ be the
message index of $W''$ respectively to be sent in block $k$. The source
encoder then transmits $x_{1k}^n(t_k|s_k,t_{k-1})$ where $t_{k-1}$ is the
index of $W''$ sent in the previous block. The relay in block $k$ will send
$x_{2k}^n (\hat t_{k-1})$, where $\hat t_{k-1}$ is the estimate of $t_{k-1}$
at the relay.

\paragraph{Decoding at relay terminal} Assume that decoding of $s_{k-1}$ and
$t_{k-1}$ in block $k-1$ has been successful. Upon receiving $y_{2k}^n$ in
block $k$, the relay looks for a unique $\hat {s}_k$ such that
\[
	\left( u_{1k}^n(\hat s_k),x_{2k}^n(\hat t_{k-1}), y_{2k}^n \right)
		\in T_{\epsilon}^n(P_{U,X_2,Y_2}).
\]
Having decoded $\hat s_k$, the relay now looks for a unique $\hat t_k$ such
that
\[
\left( x_{1k}^n(\hat{t}_k|\hat{s}_k,\hat t_{k-1}),
	u_{1k}^n(\hat s_k),x_{2k}^n(\hat{t}_{k-1}),y_{2k}^n \right)
		\in T_{\epsilon}^n(P_{U,X_1,X_2,Y_2}).
\]

\paragraph{Decoding at the sink terminal} Upon receiving $y_{3k}^n$, the
destination terminal looks for a unique $\tilde s_k$ such that
\(
\left( u_{1k}^n(\tilde{s}_k),y_{3k}^n \right) \in T_{\epsilon}^n(P_{U,Y_3})
\).
Now, the destination decodes the additional information that the source sends
in a block Markov decoding fashion. The destination terminal tries to find a
unique $\tilde{t}_{k-1}$ such that
\(
\left( x_{2k}^n(\tilde{t}_{k-1}),u_{1k}^n(\tilde s_k), y_{3k}^n \right) \in
	T_{\epsilon}^n(P_{U,X_2,Y_3})
\) and
\[
\left( x_{1k}^n(\tilde t_{k-1}|\tilde{s}_{k-1},\tilde{t}_{k-2}),
	u_{1k}^n(\tilde s_{k-1}),x_{2k}^n(\tilde t_{k-2}),y_{3(k-1)}^n
\right) \in T_{\epsilon}^n(P_{U,X_1,X_2,Y_3}).
\]

\paragraph{Rate analysis} At the relay, since we have a single user channel
from $U$ to $Y_2$, we will be able to decode the $U$ codewords with low
probability of error if $R' < I(U;Y_2|X_2)$. We can also decode the index
$t_k$ if
\[
R'' < I(X_1;Y_2|U,X_2).
\]
The destination first decodes the codeword $U$ with a low probability of error
provided $R' < I(U;Y_3)$, and then decodes the message $t_k$ using successive
interference cancellation on the messages from the relay and the source. The
message would be decoded with low probability of error provided
\[
R'' < I(X_2;Y_3|U) + I(X_1;Y_3|X_2,U).
\]
Combining all the bounds, the desired result \eqref{eq.bd} follows.
\end{IEEEproof}

In this scheme, the source message is split into two parts. The message $W'$
is broadcast to both relay and destination. And the other message $W''$ is
decoded by relay first and then cooperatively transmitted to the destination.
Unfortunately, the above achievable rate does not outperform the DF strategy,
as is shown below:
\begin{align}
R &\le \min\{ I(U;Y_2|X_2)+I(X_1;Y_2|X_2,U),
I(U;Y_3)+I(X_1X_2;Y_3|U) \} \\
& =  \min \{ I(U,X_1;Y_2|X_2), I(X_1,X_2;Y_3) \} \\
&= \min \{ I(X_1;Y_2|X_2) , I(X_1X_2;Y_3) \}.
\label{eq.DF}
\end{align}
where \eqref{eq.DF} follows from the Markov chains $U-X_1-Y_2$ and
$U-X_1-Y_3$. But \eqref{eq.DF} is the rate achieved by the Decode and Forward
strategy.

Although not providing a higher rate, the above proposed scheme of broadcast
over decode and forward gives us a good insight on the superposition strategy.
The cause of suboptimality arises due to the fact that the messages $W'$ and
$W''$ even though are generated from the same source, act as interference on
each other. This limits the rate of decoding at the relay and destination
terminals. This interference would also be present if we superimpose DF and
CF. The rate achievable using the superposition strategy is investigated in
the next section for the case of Gaussian relay channels.

\section{Achievable Rate of the Superposition Scheme} \label{sec.main}

In this section, we focus on the Gaussian relay channel. We show that when
considering only jointly Gaussian distribution for all the random variables
involved in \eqref{eq.th7}, superposition does not offer higher rate than DF
or CF alone. To be more specific, we will show that when all the random
variables involved are Gaussian, then $R_{SF}\le \max(R_{DF}, R_{CF})$.
Trivially, only one of two cases can be true
\begin{enumerate}
\item Case A: $R_{DF}\ge R_{CF}$;
\item Case B: $R_{CF}> R_{DF}$.
\end{enumerate}
It is then enough to show that in Case A, $R_{SF}\le R_{DF}$; and in Case B,
$R_{SF}\le R_{CF}$.

\subsection{Gaussian distribution assumption}

We assume that all random variables in \eqref{eq.th7} are zero mean and
jointly Gaussian distributed. The distribution will then depend only on the
variances and the cross-correlations of the random variables. For two generic
random variables $X$ and $Y$, let
\[
\phi_{X,Y}\bydef \frac{E\left\{(X-E[X])(Y-E[Y])\right\}}{\sqrt{E[X^2]E[Y^2]}}
\]
denote the correlation coefficient between them. The following lemma is useful
in deducing correlations from known ones.

\begin{lemma} \label{lem.xc}
Let $X-Y-Z$ be a Markov chain of jointly Gaussian random variables. Then
$\corr XZ = \corr XY \corr YZ$.
\end{lemma}
\begin{IEEEproof}
See appendix.
\end{IEEEproof}

\begin{figure}
\centering
\includegraphics[width=2in]{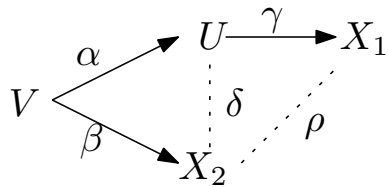}
\caption{Dependency graph of random variables with correlation coefficients}
\label{correlationfig}
\end{figure}

Returning to the random variables involved in $R_{SF}$, we denote
$\alpha=\corr UV$, $\beta=\corr V{X_2}$, and $\gamma=\corr U{X_1}$. Using
\lemref{lem.xc}, we obtain from the Markov chain $U-V-X_2$ that
\begin{align}
\delta &\bydef \corr {X_1}{X_2} = \corr VU\cdot \corr V{X_2} = \alpha \beta, \\
\intertext{and from the Markov chain $X_1-U-X_2$ that}
\rho &\bydef \corr {X_1}{X_2} = \corr {X_1}U \cdot \corr U{X_2} = \gamma \delta
=\alpha\beta\gamma.
\end{align}
\figref{correlationfig} shows the correlation between the random variables
along with their dependencies on each other.

\subsection{Main Result}

The main result is stated in the following theorem. Two lemmas that are needed
in the proof are stated and proved in the appendix.
\begin{theorem}\label{th.main}
Let $(X_1,X_2,Y_2,Y_3,\hat{Y_2'},\hat{Y_2},U,V,)$ be a set of jointly Gaussian
random variable whose joint distribution can be factorized in the following
form:
\begin{equation}\label{eq.probdist}
p(u,v,x_1,x_2,y_2,y_3,\hat y_2') = p(v)p(u|v)p(x_2|v)p(x_1|u)
p(y_2,y_3|x_1x_2)p(\hat y_2'|y_2,u,x_2)p(\hat y_2|y_2, x_2),
\end{equation}
where $p(y_2,y_3|x_1x_2)$ is as given in \eqref{eq.ch}. Let $\cP$ denote the
class of distributions specified by \eqref{eq.probdist}. Let $\cP'$ denote a
subset of $\cP$ with distributions that also satisfy the constraint
\eqref{eq.th7con}. We have
\begin{align}
\sup_{\cP'}  \min \{ &I(X_1;Y_3,\hat Y_2'|X_2,U)+I(U;Y_2|X_2,V),
	I(X_1,X_2;Y_3)-I(\hat Y_2';Y_2|X_1,X_2,U,Y_3)\} \label{eq.SF} \\
	= \max \{ &\sup_\cP \min \{ I(X_1;Y_2|X_2),I(X_1,X_2;Y_3) \},
		\label{eq.RDF}\\
 & \sup_\cP \min \{ I(X_1;\hat{Y}_2,Y_3|X_2),
 	I(X_1,X_2;Y_3)- I(\hat{Y}_2;Y_2|X_1,X_2,Y_3) \} \}.\label{eq.CF}
\end{align}
\end{theorem}

\begin{IEEEproof}
The rates appearing in \eqref{eq.SF}--\eqref{eq.CF} are $R_{SF}$, $R_{DF}$,
and $R_{CF}$, respectively. Since through the judicious choice the random
variables $U$ and $V$, DF and CF can be cast as special cases of SF
\cite{coga79}, we have $R_{SF}\ge R_{DF}$ and $R_{SF}\ge R_{CF}$. It is then
sufficient to show that $R_{SF}\le \max(R_{DF}, R_{CF})$.

Under the Gaussian assumption, the compressed version $\hat Y_2'$ of $Y_2$ in
\eqref{eq.p7} can be written as
\begin{equation}  \label{eq.aux}
\hat{Y}_2' = c_1 Y_2 + c_2 U + c_3 X_2 + Z_w'
\end{equation}
where $c_1, c_2, c_3$ are constant parameters, $Z_w'$ is Gaussian and
independent of $Y_2$, $U$, and $X_2$. Since in both \eqref{eq.th7} and
\eqref{eq.th7con}, the three mutual information terms involving $\hat {Y}_2'$,
namely,
\[
I(X_1; Y_3, \hat {Y}_2' | X_2,U),\quad
I(\hat {Y}_2'; Y_2|X_2, X_1, U, Y_3), \quad
I(\hat {Y}_2'; Y_2|X_2, U, Y_3)
\]
are all conditioned on $U$ and $X_2$, the coefficients $c_2$ and $c_3$ do not
affect the values of these terms. Therefore we can set $c_2=c_3=0$. It is also
true that scaling $\hat {Y}_2'$ by a constant does not change any of the
terms. So unless $c_1=0$, we can assume $c_1=1$, as we do in the following.
The case $c_1=0$ is known as the so called partial decoding and forward
scheme, which is known to be inferior to the full DF scheme \cite{elmz06}. We
denote the variance of $Z_w'$ as $\Delta'$. The amount of compression, which
is controlled by the parameter $\Delta'$, depends on the constraint
\eqref{eq.th7con} imposed by the relay link channel and the encoding scheme at
the relay. In summary, we can take without loss of generality
\begin{equation} \label{eq.y2hat'}
	\hat {Y}_2'=Y_2 + Z_w',
\end{equation}

The following is a broad outline of the proof. Given any rate achieved by the
SF scheme, we can find a CF scheme or a DF scheme which can achieve a rate
higher than or equal to SF. The $\hat{Y}_2$ for the CF scheme is set to be
statistically equal to $\hat {Y}_2'$ of the SF scheme in \eqref{eq.y2hat'}:
\begin{equation} \label{eq.y2hat}
        \hat {Y}_2=Y_2 + Z_w,
\end{equation}
where $Z_w$ is zero mean Gaussian with variance $\Delta=\Delta'$. Such $\hat
Y_2$ would qualify as the compressed version of $Y_2$ in CF. This choice of
$\hat{Y}_2$ is enough to achieve a higher rate than SF even though it can be
suboptimal to the possible rates achievable by CF.

First, we have
\begin{align}
I(\hat Y_2'; Y_2|X_1, X_2, U, Y_3)
	&=h(Y_2|X_1, X_2, U, Y_3)-h(Y_2|X_1, X_2, U, Y_3, \hat Y_2')\label{eq.c1}\\
	&= h(Y_2|X_1, X_2, Y_3)-h(Y_2|X_1, X_2, U, Y_3, \hat Y_2')\label{eq.c2}\\
	&\ge h(Y_2|X_1, X_2, Y_3)-h(Y_2|X_1, X_2, Y_3, \hat Y_2')\label{eq.c3}\\
	&\ge h(Y_2|X_1, X_2, Y_3)-h(Y_2|X_1, X_2, Y_3, \hat Y_2)\label{eq.c4}\\
	&= I(\hat Y_2; Y_2|X_1, X_2, Y_3)
\end{align}
where \eqref{eq.c2} is due to the Markov chain $U-(X_1, X_2, Y_3)-Y_2$;
\eqref{eq.c3} uses the fact that conditioning does not increase entropy; and
\eqref{eq.c4} is because given $(X_2, U)$, $\hat Y_2'$ is statistically
equivalent to $\hat Y_2$.

Thus, we have shown
\begin{equation}
	I(X_1,X_2;Y_3)-I(\hat Y_2';Y_2|X_1,X_2,U,Y_3)
		\leq I(X_1,X_2;Y_3)-I(\hat Y_2;Y_2|X_1,X_2,Y_3).
\end{equation}
It then remains to be shown that
\begin{equation} \label{eq.main}
	I(X_1;Y_3,\hat Y_2'|X_2,U)+I(U;Y_2|X_2,V)\le
		\max \{ I(X_1;Y_2|X_2), I(X_1;\hat{Y}_2,Y_3|X_2) \}.
\end{equation}
Depending on which one of the two terms on the right hand side is bigger, we
have two cases. In the first case,
\begin{equation} \label{eq.cond1}
	I(X_1;Y_2|X_2) \ge I(X_1;Y_3,\hat{Y_2}|X_2)
\end{equation}
and we have
\begin{align}
&\quad\;\, I(U;Y_2|V,X_2) + I(X_1;Y_3,\hat{Y_2'}|X_2,U) \\
&= I(U;Y_2|V,X_2) + I(X_1;Y_3,\hat{Y_2}|X_2,U) \label{eq.2.5}\\
&=  I(U;Y_2|X_2) - I(V;Y_2|X_2) + I(X_1;Y_3,\hat{Y_2}|X_2,U) \label{eq.3}\\
&=  I(X_1;Y_2|X_2) - I(X_1;Y_2|X_2,U)
 - I(V;Y_2|X_2) + I(X_1;Y_3,\hat{Y_2}|X_2,U) \label{eq.4}\\
& \le  I(X_1;Y_2|X_2) - I(X_1;\hat{Y_2},Y_3|X_2,U)
 - I(V;Y_2|X_2) + I(X_1;Y_3,\hat{Y_2}|X_2,U) \label{eq.5}\\
& =  I(X_1;Y_2|X_2) - I(V;Y_2|X_2) \label{eq.6}\\
& \le  I(X_1;Y_2|X_2) \label{eq.7}
\end{align}
where \eqref{eq.2.5} follows by our choice of $\hat Y_2$ to be statistically
the same as $\hat Y_2'$; \eqref{eq.3} follows from the Markov chain $V-(U,
X_2)-Y_2$; \eqref{eq.4} follows from the Markov chain $U-(X_1,X_2)-Y_2$;
\eqref{eq.5} follows from \eqref{eq.cond1} and Lemma 2, which is stated and
proved in \apref{sec.lemmas}; and \eqref{eq.7} follows from the fact that
mutual information is nonnegative.

In the second case,
\begin{equation} \label{eq.cond2}
	I(X_1;Y_2|X_2) < I(X_1;Y_3,\hat{Y_2}|X_2)
\end{equation}
and we have
\begin{align}
&\quad\;\, I(X_1;Y_3,\hat{Y}_2'|X_2,U) + I(U;Y_2|V,X_2) \nonumber\\
&= I(X_1;Y_3,\hat{Y}_2|X_2,U) + I(U;Y_2|V,X_2) \label{eq.thr0} \\
&= I(X_1;Y_3,\hat{Y}_2|X_2,U) + I(U;Y_2|X_2) - I(V;Y_2|X_2) \label{eq.thr1} \\
&\le I(X_1;Y_3,\hat{Y}_2|X_2,U) +
	I(U;Y_3,\hat{Y}_2|X_2) -I(V;Y_2|X_2) \label{eq.thr2} \\
&= I(X_1;Y_3,\hat{Y}_2|X_2) - I(V;Y_2|X_2) \label{eq.thr3} \\
&\le I(X_1;Y_3,\hat{Y}_2|X_2) \label{eq.thr4}
\end{align}
where \eqref{eq.thr0} follows by our choice of $\hat Y_2$ to be statistically
the same as $\hat Y_2'$; \eqref{eq.thr1} follows from the Markov chain
$V-(U,X_2)-Y_2$; \eqref{eq.thr2} follows from \eqref{eq.cond2} and Lemma 3,
which is stated and proved in \apref{sec.lemmas}; \eqref{eq.thr3} follows from
the Markov chain $U - (X_1, X_2) - \hat{Y_2},Y_3$; and \eqref{eq.thr4} follows
from the fact that mutual information is nonnegative.

Thus we have shown \eqref{eq.main} holds. And the whole proof is complete.
\end{IEEEproof}

\subsection{Discussion}

We have shown that the SF does not outperform both DF and CF. We provide some
intuitive explanation in the following.

Observe from \eqref{eq.y2hat} that $\hat{Y_2}$ is the quantized signal of
$Y_2$ in the CF scheme. The variance of $Z_w$ is $\Delta$, which in general
could be different from $\Delta'$, the variance of $Z_w'$ in \eqref{eq.aux}.
From the constraint \eqref{eq.RCF}, we have $\Delta \geq \Delta_{CF}$, where
\begin{equation}
\Delta_{CF} = \frac{N_1N_2 + \left(N_1 + a^2 N_2 \right)P_1}{b^2 P_2}.
\end{equation}
Although the constraint is not explicitly imposed in the formulation in
\eqref{eq.CF2}, it can be shown that setting $\Delta=\Delta_{CF}$ actually
maximizes the two terms on the right hand side of \eqref{eq.CF2}, and
equalizes them:
\begin{equation}
	I(X_1;\hat{Y}_2,Y_3|X_2)=I(X_1,X_2;Y_3)- I(\hat{Y}_2;Y_2|X_1,X_2,Y_3).
\end{equation}
It can be verified that
\begin{enumerate}
\item $I((X_1;\hat{Y}_2,Y_3|X_2)$ is a monotonically decreasing function of
$\Delta$ (coarser compression reduces the useful information about $X_1$ in
$\hat Y_2$);
\item $I(\hat{Y}_2;Y_2|X_1,X_2,Y_3)$ is a monotonically increasing function of
$\Delta$.
\end{enumerate}
Therefore the minimum of the two functions is maximized at their crossing
point, which happens at $\Delta=\Delta_{CF}$. In other words, for CF, within
the relay-destination link rate limit $I(X_2; Y_3)$, more compression yields
higher rate over all. For the SF, however, the situation is different. The
parameter $\Delta'$, which controls the amount of compression in
\eqref{eq.aux} needs to be chosen to satisfy the constraint \eqref{eq.th7con}.
In particular, we have $\Delta' \geq \Delta_{SF}$, where
\begin{equation}\label{eq.DeltaSF}
	\Delta_{SF} =
	\frac{(N_2 + P_1(1-\alpha^2\gamma^2))
		(N_1N_2 + (N_1+a^2 N_2)P_1(1-\gamma^2))}
	{b^2P_2(1-\beta^2)[N_2+P_1(1-\gamma^2)]}
\end{equation}
In general $\Delta_{SF}$ can be less than $\Delta_{CF}$; e.g., when $\gamma >
0$, $\alpha =1$ and $\beta=0$. In contrast to the CF case, it is not true for
SF that more compression (smaller $\Delta'$) necessarily yields higher rate.
The intuitive reason is that the relay has two messages to transmit to the
destination: the partially decoded message carried by $U$ and the compressed
version of $Y_2$ carried by $\hat Y_2'$. Although reducing $\Delta'$ will
provide to the destination a more faithful representation of $Y_2$, and
enlarge the term $I(X_1; Y_3, \hat Y_2'|X_2,U)+I(U;Y_2|X_2,V)$, it will reduce
the relay's ability to cooperate with the source through the message $U$, and
hence enlarge the gap $I(\hat Y_2';Y_2|X_1,X_2,U,Y_3)$ from the
multiple-access cut-set bound $I(X_1, X_2; Y_3)$, which then becomes the rate
limiting factor. The optimum amount compression turns out to the be same as in
the CF case. And superposition of DF and CF does not help the rate, which
agrees with the observation that we have made in \secref{sec.super}.

Finally, we remark that in our proof we did not use the constraint
\eqref{eq.th7con}. So it is true that for the Gaussian distribution, even
without the constraint, the SF does not result in a rate that is higher than
the larger one of $R_{DF}$ and $R_{CF}$.

\section{Numerical Result} \label{sec.num}

Considering an example Gaussian relay channel such that the source and the
destination are separated by a unit distance, and the relay is at distance $d$
from the source and $1-d$ from the destination. The channel gain between any
two nodes is inversely proportional to their distance. So $a=1/d$ and
$b=1/(1-d)$. The additive noises at the relay and the destination are
independent but have the same variance $N_1=N_2=1$. The transmit powers are
set to $P_1=P_2=5$.

\begin{figure}[t]
\centering
\includegraphics[width=\linewidth]{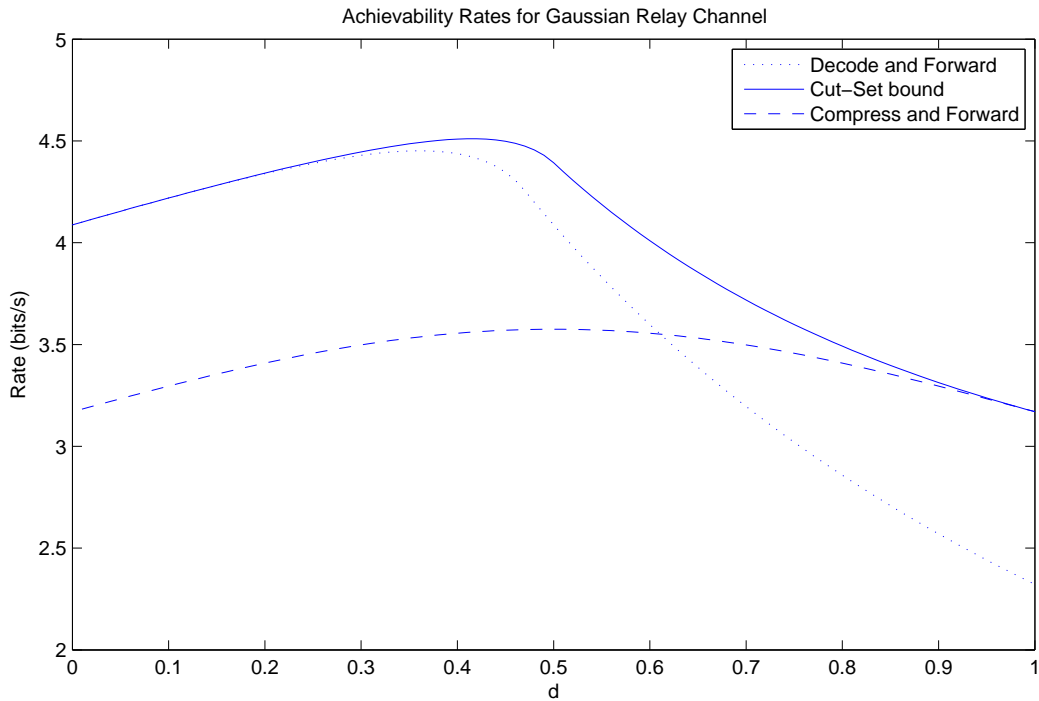}
\caption{Achievable rates for the Gaussian relay channel, where $d$ is the
normalized distance from source to relay.}
\label{fig.DFCF}
\end{figure}

\figref{fig.DFCF} shows the numerical rates achievable by DF, CF and the
cutset bound \eqref{eq.cutset} as a function of distance $d$ of the relay from
the source terminal. Depending on $d$, there are three cases:
\begin{enumerate}
\item When $d$ is small (roughly $d<0.2$), DF is optimal. The rate achieved by
$DF$ is equal to $I(X_1, X_2; Y_3)$ the multiple-access cut-set bound. The
reason is that the source message can be fully decoded at the relay.

\item For medium $d$ (roughly $0.2<d<0.6$), DF is not optimal, but still
performs better than CF. In this case, the rate of DF is dominated by $I(X_1,
Y_2|X_2)$, the amount information can be decoded at the relay, which dictates
the amount of cooperation possible between source and relay. In this region,
the relay-sink channel is ``poor'' so that sending ``finely'' compressed
version of $Y_2$ is not possible.

\item For large $d$ (roughly $0.6<d\le 1$), CF out performs DF. In this
region, the ability of the relay to decode the source is weak, and it is more
fruitful to send compressed version of the relay's observation. Only in the
extreme case, $d=1$, does CF actually achieve the cut-set bound.
\end{enumerate}

\begin{figure}
\centering
\includegraphics[width=\linewidth]{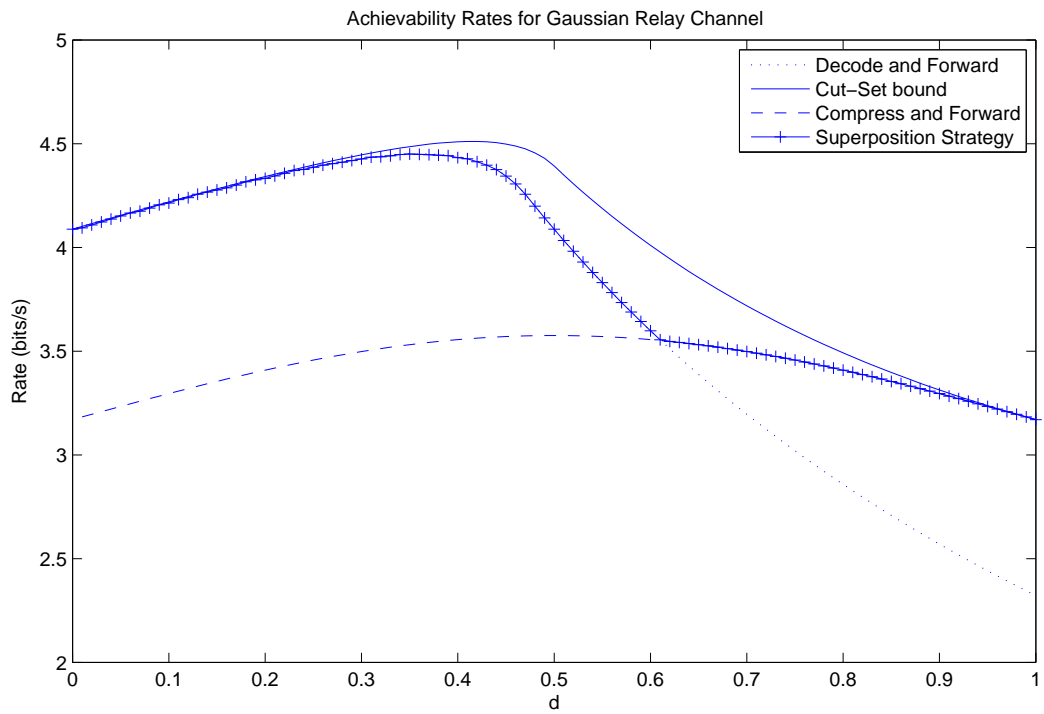}
\caption{Achievable rates for Gaussian relay channel. The parameters of the
superimposing strategy are optimized to maximize the achievable rate}
\label{fig.SF}
\end{figure}

The rate achievable by superimposing DF and CF given by \eqref{eq.th7} is
numerically compared with the rates achieved by CF, DF and the cut-set bound.
The mutual information terms of \eqref{eq.th7} are evaluated for the choice of
appropriate Gaussian Random variables, according to \eqref{eq.g4} and
\begin{align}
&I(U;Y_2|X_2,V) = C \left(\frac{\frac{P_1}{d^2} \gamma^2 (1- \alpha^2) }
	{N_1 + \frac{P_1}{d^2}(1-\gamma^2)} \right), \\
&I(X_1X_2;Y_3) = C \left( \frac{P_1+\frac{P_2}{(1-d)^2}
	+ \frac{2 \rho \sqrt{P_1P_2}}{(1-d)}}{N_2} \right), \\
&I(Y_2;\hat{Y}_2|U,X_1,X_2,Y_3) = C \left( \frac{N_1}{\Delta}\right).
\end{align}
The constraint $I(\hat{Y}_2;Y_2|U,X_2,Y_3) \le I(X_2;Y_3|V) $ is evaluated to
$\Delta'\ge \Delta_{SF}$, where $\Delta_{SF}$ is as given in
\eqref{eq.DeltaSF}. The correlation terms $\alpha, \beta, \gamma$ and the
variance $\Delta'$ are optimizing parameters, which control the amount of
information that is decoded and the amount that is compressed. When all the
parameters have been optimized within the constraint posed by
\eqref{eq.DeltaSF}, the SF is found to achieve the maximum of $R_{DF}$ and
$R_{CF}$, as shown in \figref{fig.SF}.

\section{Conclusion} \label{sec.conc}

We analyzed the coding strategy of superimposing CF and DF for the Gaussian
relay channel. We note that superposition of CF and DF does not provide higher
achievable rates than the individual DF and CF for the Gaussian case. We
conclude that we should look for new strategies different from superposition
strategy, or look for non-Gaussian distributions for the superposition scheme,
or try to find tighter upper bounds than the cut-set bound.

\appendix

\subsection{Proof of \lemref{lem.xc}}
\begin{IEEEproof}
Assume without loss of generality that all three random variables are zero
mean. We have
\begin{equation}
\begin{split}
\corr XZ &= \frac{\E [XZ]}{\sqrt{\E[X^2]\E[Z^2]}} \\
	&= \frac{\E\{\E[XZ|Y]\}}{\sqrt{\E[X^2]\E[Z^2]}}\\
	&= \frac{\E\{\E[X|Y]\E[Z|Y]\}}{\sqrt{\E[X^2]\E[Z^2]}}\\
	&= \frac{\E\{\sqrt{\E[X^2]/\E[Y^2]}\corr XY Y \cdot
		\sqrt{\E[Z^2]/\E[Y^2]}\corr YZ Y \} } {\sqrt{\E[X^2]\E[Z^2]}}\\
	&=\corr XY \corr YZ
\end{split}
\end{equation}
\end{IEEEproof}

\subsection{Two lemmas needed in the proof of \thref{th.main}}
\label{sec.lemmas}

We prove two lemmas in the following that will be useful in the proof of
\thref{th.main}. \lemref{lem.2} is used in the case $R_{DF}\ge R_{CF}$.
\lemref{lem.3} is used in the case $R_{DF}<R_{CF}$.

\begin{lemma}\label{lem.2}
Let $(X_1,X_2,Y_2,Y_3,\hat{Y_2},U,V) $ be jointly Gaussian random variables
with joint distribution
\(
	p(u,v,x_1,x_2,y_2,y_3,\hat{y_2}) =
		p(v)p(u|v)p(x_2|v)p(x_1|u)p(y_2,y_3|x_1x_2)p(\hat{y_2}|y_2,x_2)
\),
where $p(y_2,y_3|x_1,x_2)$ is as given in \eqref{eq.ch}. If $I(X_1;Y_2|X_2)
\ge I(X_1;Y_3,\hat{Y_2}|X_2)$ then $I(X_1;Y_2|X_2,U) \ge
I(X_1;\hat{Y_2},Y_3|X_2,U)$.
\end{lemma}
\begin{IEEEproof}
Under the Gaussian assumption, we have
\begin{align}
&I(X_1;Y_2|X_2)  = \frac{1}{2} \log \left\{ 1 + \frac{a^2 P_1(1 - \rho^2) }{N_1}  \right\} \label{eq.g1} \\
&I(X_1;Y_2|X_2,U) = \frac{1}{2} \log \left\{ 1 + \frac{a^2 P_1(1 - \gamma^2)}{N_1}\right\} \label{eq.g2} \\
&I(X_1;\hat{Y}_2,Y_3|X_2) =  \frac{1}{2} \log \left\{ 1 + P_1(1-\rho^2) \frac{(N_1+\Delta)+ a^2 N_2}{(N_1+\Delta)N_2}  \right\}  \label{eq.g3} \\
&I(X_1;\hat{Y}_2,Y_3|X_2,U)= \frac{1}{2} \log \left\{ 1 + P_1(1-\gamma^2) \frac{(N_1+\Delta) + a^2 N_2}{(N_1+\Delta)N_2} \right\} \label{eq.g4}
\end{align}

Obviously when $\rho=1$ and hence $\gamma=1$ (because
$\rho=\alpha\beta\gamma$), the lemma holds. We thus assume that $\rho<1$.
Since $I(X_1;Y_2|X_2) \ge I(X_1;\hat{Y}_2,Y_3|X_2)$, from \eqref{eq.g1} and
\eqref{eq.g2} we have
\begin{equation}
	\frac{a^2 P_1(1 - \rho^2)}{N_1} \ge P_1(1-\rho^2)
		\frac{(N_1+\Delta) + a^2 N_2}
			{(N_1+\Delta)N_2}.
\end{equation}
Multiplying both sides with $(1-\gamma^2)/(1-\rho^2)$, we obtain
\begin{equation}
	\frac{a^2 P_1(1 - \gamma^2) }{N_1}  \ge P_1(1-\gamma^2)
		\frac{(N_1+\Delta)+ a^2 N_2}
			{(N_1+\Delta)N_2}.
\end{equation}
It then follows that $I(X_1;Y_2|X_2,U) \ge I(X_1;\hat{Y}_2,Y_3|X_2,U)$ from
the monotonic property of the logarithmic function.
\end{IEEEproof}

\begin{lemma}\label{lem.3}
\noindent Let $(X_1,X_2,Y_2,Y_3,\hat{Y_2},U,V) $ be jointly Gaussian random
variables with distribution
\(
p(u,v,x_1,x_2,y_2,y_3,\hat{y_2}) = p(v)p(u|v)p(x_2|v)p(x_1|u)
p(y_2,y_3|x_1,x_2)p(\hat{y_2}|y_2,x_2) \nonumber
\), where $p(y_2,y_3|x_1,x_2)$ is as given in \eqref{eq.ch}.
If $I(X_1;Y_2|X_2) \le I(X_1;\hat{Y}_2,Y_3|X_2)$ then $I(U;Y_2|X_2) \le
I(U;\hat{Y}_2,Y_3|X_2)$.
\end{lemma}

\begin{IEEEproof}
Under the Gaussian variable assumptions, we have
\begin{align}
&I(X_1;Y_2|X_2)  = \frac{1}{2} \log \left\{ 1 +
	\frac{a^2 P_1(1 - \rho^2) }{N_1}  \right\} \label{eq.r1} \\
&I(U;Y_2|X_2) = \frac{1}{2} \log \left\{ 1 +
	\frac{a^2 P_1(\gamma^2 - \rho^2)}
		{N_1 + a^2 P_1(1- \gamma^2) }\right\} \label{eq.r2} \\
&I(X_1;\hat{Y}_2,Y_3|X_2) =  \frac{1}{2} \log \left\{ 1 +
	P_1(1-\rho^2) \frac{(N_1+\Delta)+ a^2 N_2}
		{(N_1+\Delta)N_2}  \right\}  \label{eq.r3} \\
&I(U;\hat{Y}_2,Y_3|X_2)= \frac{1}{2} \log \left\{ 1 +
	\frac{P_1(\gamma^2 - \rho^2)[(N_1+\Delta) + a^2 N_2]}
		{(N_1+\Delta)N_2 + P_1(1-\gamma^2)[(N_1+\Delta) + a^2 N_2] }
	\right\} \label{eq.r4}
\end{align}

It can be verified that when $\gamma=1$, $I(X_1;Y_2|X_2)=I(U;Y_2|X_2)$ and
$I(X_1;\hat{Y}_2,Y_3|X_2)=I(U;\hat{Y}_2,Y_3|X_2)$, so that the desired result
holds in this case. In the following, we assume that $\gamma<1$, and therefore
$\rho=\alpha\beta\gamma<1$.

Since $I(X_1;Y_2|X_2) \le I(X_1;\hat{Y}_2,Y_3|X_2)$, it follows from
\eqref{eq.r1} and \eqref{eq.r2} that
\begin{equation}\label{eq.p1}
	\frac{a^2 P_1(1 - \rho^2) }{N_1} \le
		\frac{P_1(1-\rho^2)[(N_1+\Delta)+ a^2 N_2]}{(N_1+\Delta)N_2}.
\end{equation}
Multiplying both sides of \eqref{eq.p1} with $(1-\gamma^2)/(1-\rho^2)$ we
obtain
\begin{equation}\label{eq.p2}
	\frac{a^2 P_1(1 - \gamma^2) }{N_1} \le
		\frac{P_1(1-\gamma^2) [(N_1+\Delta)+ a^2 N_2]}{(N_1+\Delta)N_2}.
\end{equation}
Adding the numerator to the denominator on both sides, we obtain
\begin{equation}\label{eq.p3}
	\frac{a^2 P_1(1 - \gamma^2) }{N_1+a^2 P_1(1 - \gamma^2)} \le
		\frac{P_1(1-\gamma^2)[(N_1+\Delta)+ a^2 N_2]}
			{(N_1+\Delta)N_2+P_1(1-\gamma^2)[(N_1+\Delta)+ a^2 N_2]}.
\end{equation}
Multiplying both sides of \eqref{eq.p3} by $(\gamma^2-\rho^2)/(1-\gamma^2)$,
we obtain
\begin{equation}\label{eq.p4}
	\frac{a^2 P_1(\gamma^2 - \rho^2) }{N_1+a^2 P_1(1 - \gamma^2)} \le
		\frac{P_1(\gamma^2-\rho^2)[(N_1+\Delta)+ a^2 N_2]}
			{(N_1+\Delta)N_2+P_1(1-\gamma^2)[(N_1+\Delta)+ a^2 N_2]}
\end{equation}
It then follows that $I(U;Y_2|X_2) \le I(U;\hat{Y}_2,Y_3|X_2)$ due to the
monotonic property of the logarithmic function.
\end{IEEEproof}

\bibliographystyle{IEEEtran}
\bibliography{refs}

\begin{thebibliography}{1}
\providecommand{\url}[1]{#1}
\def\UrlFont{\rmfamily}
\providecommand{\newblock}{\relax}
\providecommand{\bibinfo}[2]{#2}
\providecommand\BIBentrySTDinterwordspacing{\spaceskip=0pt\relax}
\providecommand\BIBentryALTinterwordstretchfactor{4}
\providecommand\BIBentryALTinterwordspacing{\spaceskip=\fontdimen2\font plus
\BIBentryALTinterwordstretchfactor\fontdimen3\font minus
  \fontdimen4\font\relax}
\providecommand\BIBforeignlanguage[2]{{%
\expandafter\ifx\csname l@#1\endcsname\relax
\typeout{** WARNING: IEEEtran.bst: No hyphenation pattern has been}%
\typeout{** loaded for the language `#1'. Using the pattern for}%
\typeout{** the default language instead.}%
\else
\language=\csname l@#1\endcsname
\fi
#2}}

\bibitem{vand71}
E.~C. van~der Meulen, ``Three-terminal communication channels,'' \emph{Advanced
  Applied Probability}, vol.~3, pp. 120--154, 1971.

\bibitem{coga79}
T.~Cover and A.~El~Gamal, ``Capacity theorems for the relay channel,''
  \emph{{IEEE} Trans. Inf. Theory}, vol.~25, no.~5, pp. 572--584, May 1979.

\bibitem{gaar82}
\BIBentryALTinterwordspacing
A.~E. Gamal and M.~Aref, ``The capacity of the semideterministic relay
  channel,'' \emph{{IEEE} Trans. Inform. Theory}, vol.~28, no.~3, pp. 536--536,
  Mar. 1982. [Online]. Available: \url{doi:10.1109/TIT.1982.1056502}
\BIBentrySTDinterwordspacing

\bibitem{elmz06}
A.~El~Gamal, M.~Mohseni, and S.~Zahedi, ``Bounds on capacity and minimum
  energy-per-bit for {AWGN} relay channels,'' \emph{{IEEE} Trans. Inf. Theory},
  vol.~52, no.~4, pp. 1545--1561, Apr. 2006.

\end{thebibliography}

\clearpage

\end{document}